# Enhanced density fluctuations in water-ethanol mixtures at low ethanol concentrations: Temperature dependent studies

**Rikhia Ghosh, Biman Bagchi***

E-mail: bbagchi@sscu.iisc.ernet.in

Solid State and Structural Chemistry Unit, Indian Institute of Science, Bangalore 560012, India

## *Abstract*

Since the structural transformations observed in water-ethanol binary mixtures are apparently driven by relatively weak intermolecular forces (like hydrophobicity and hydrogen bonding) that often cooperate to form self-assembled structures, one expects the aggregation properties to show strong temperature dependence. We study the effect of temperature on the formation of transient ethanol clusters as well as on the dynamic density heterogeneity induced in the system due to such clustering. The dynamic heterogeneity is expected to occur on small length scales with short lifetime – both are expected to be temperature dependent. Indeed, a major finding of the work is strong temperature dependence of the extent of structural heterogeneity. Distinct signature of static and dynamic heterogeneity of ethanol molecules is also found to appear with lowering of temperature. This is attributed to the formation of transient ethanol clusters that are known to exhibit considerably small lifetime (order of a few picosecond). The transient dynamical features of dynamic heterogeneity are expected to affect those relaxation processes occurring at sub-picosecond time scales. On the other hand, strong temperature dependence of micro-structure formation can be anticipated to be due to enhanced structural order stimulated in the system with lowering of temperature. Present analyses reveal a number of interesting features which were not explored beforehand in this widely studied binary mixture.

# I. Introduction

Aqueous binary mixtures are known to often exhibit exotic composition dependent thermodynamic and dynamic properties, observed in both experiments and simulations[1]. These non-trivial behaviors are largely interpreted in terms of structural transformations driven by a combination of contrasting intermolecular interactions (hydrophobic interactions, hydrogen bonding). Such intermolecular interactions cooperatively determine local arrangements and also an underlying free energy surface which can be quite rugged (on small energy scale) because of the possibility of many close lying minima reflecting different molecular arrangements. However detailed microscopic understanding of precise nature of these structural arrangements and transformations among them has still remained somewhat vague. Since the observed structural transformations are anticipated to be majorly driven by relatively weak intermolecular forces, one can expect these properties to show substantial temperature and pressure dependence. In this context, temperature and pressure dependent studies of such aqueous binary mixture systems are expected to offer valuable insights about the transient structural heterogeneity persistent in the systems. In particular, the structural transformations are anticipated to be severely affected by thermal effects. In this work, we explore such possibility by studying the temperature dependence of structural transformations and related changes in structural and dynamic properties in aqueous ethanol mixtures.

Series of experiments have revealed a host of striking anomalies in water-ethanol binary mixture over a wide range of composition. Several thermodynamic and transport properties, including excess entropy, molar volume, diffusion coefficient, compressibility, viscosity, Walden product, sound attenuation coefficient, etc.[2-7], show significant non-ideal deviations. In most cases, the concentration dependence of these thermodynamic properties is found to show either maxima or minima in the low concentration region. The isentropic compressibility shows well-defined minimum at $x_{EtOH}$ = 0.08. The excess enthalpy of mixing also shows a minimum at $x_{EtOH}$ = 0.12 at 25°C[4]. On the other hand, the partial molar volumes indicate that the co-solvent apparently contracts up to an ethanol concentration of about a mole fraction of ethanol, $x_{EtOH}$ = 0.08[2]. Frank and Evans[3] first promoted the idea of formation of a low entropy cage of water with stronger H-bonds in water-alcohol systems, popularly known as the "iceberg" model in order to explain the composition-dependent anomalies. In spite of a broad support for this view[8,9], a different perspective is suggested by recent scattering experiments[10,11]. Soper et al.[12,13] found that there are only minor changes in hydrogen bond occurrence in the first hydration shell of the solute (alcohol), but the major structural change happens in the second hydration shell. Formation of compact structure in the second hydration shell was then considered to be responsible for the anomalous behaviors in the dilute alcohol-water solutions.

A series of experiments have shown existence of distinct structural regimes in water-ethanol binary mixture. Differential scanning calorimetry studies[14] suggested four regimes – the transition point between the first two regimes was around $x_{EtOH}$ = 0.12 while the other transition points were found at $x_{EtOH}$ = 0.65 and 0.85. These findings were in agreement with earlier NMR and Fourier transform infrared studies[15]. Nishi and co-workers[9] explored through low frequency Raman spectroscopy a change in local structure at $x_{EtOH}$ = 0.20. This was supposedly due to two separate states of the system; ethanol aggregated state and the water aggregated state. Interestingly, they suggested that the interactions between the ethanol aggregates and the water aggregates are weak to lead microscopic phase separation. Mass spectrometric techniques have been used a number of times[16-19] to understand the structure of water-ethanol binary mixture.

Nishi et al.[16] found the presence of $(C_2H_5OH)_m(H_2O)_n$ species only below $x_{EtOH}< 0.04$. Above $x_{EtOH}\sim 0.04$, they observed ethanol aggregates, which they termed "polymers" of ethanol. At $x_{EtOH}= 0.08$, they found that the growth of ethanol "polymers" is almost saturated. Surprisingly, the intensity of the "polymers" became weaker with increasing ethanol concentration beyond $x_{EtOH} = 0.42$, and neat ethanol do not show any aggregation. In a mass spectrometric study done later, Wakisaka et al.[19] demonstrated that the ethanol-water binary mixtures have microscopic phase separation at the cluster level beyond $x_{EtOH} = 0.03$. Biswas and co-workers[20] studied the absorption and emission spectrum of coumarin 153 in water-ethanol binary mixture. They observed maximum in the peak frequencies at $x_{EtOH} = 0.10$ and 0.20 respectively for absorption and emission – which are believed to be due to structural changes in the system. Raman studies[21] on stretching bands pointed to a structural rearrangement at $x_{EtOH} = 0.05$-0.10. Juurinen et al.[22] employed x-ray Compton scattering to investigate the intra- and inter-molecular bond lengths in ethanol-water mixtures. They found that at low ethanol concentration ($x_{EtOH}< 0.05$) all the O–H covalent bonds (for both water and ethanol) are elongated which corresponds to strong inter-molecular hydrogen bonds. At high ethanol concentration ($x_{EtOH} = 0.15$-0.73), the inter-molecular hydrogen bonds contract markedly leading to an increase in density. This study indicates that a structural re-arrangement of ethanol-water mixture occurs between $x_{EtOH} = 0.06$-0.15. A recent work by Perera and coworkers[23] further highlights the scenario of micro-heterogeneity in aqueous ethanol solution. By means of ultrasonic and hypersonic measurements and molecular dynamics simulation they show that these mixtures show aggregation of ethanol molecules in the low ethanol mole fraction $x_{EtOH}< 0.2$, bicontinuous-like phase around $x_{EtOH} = 0.5$, and weak water clustering above $x_{EtOH} = 0.8$.

The prior computational studies of water-ethanol binary mixtures have been mostly targeted toward reproducing the properties and anomalies at different concentrations[24-26], or understanding the hydrophobic hydration[27]. Fidler and Rodger[28] used molecular dynamics simulation to characterize the structure of water around ethanol. The static structure of water around the hydrophobic end of the alcohol was found to be essentially the same as that found in bulk water. Khattab et al. have reported the composition dependent measured value of density, viscosity and surface tension of water-ethanol binary mixture at a number of different temperatures and compared with available literature data[29]. We showed in a prior work that there is an abrupt emergence of a bi-continuous phase at low ethanol concentration ($x_{EtOH}$ = 0.06-0.1) that is attributed to a percolation-like phase transition[30]. We also showed that the collapsed state of a linear homopolymer chain gains surprising stability at low ethanol concentration ($x_{EtOH}$ = 0.05) that is expected to be an outcome of micro-heterogeneous phase separation of aqueous ethanol solution at low concentration.

In this context, it is very important to look into the solid-liquid phase diagram of water-ethanol binary mixture. The phase equilibrium of water-ethanol system is quite complicated due to the existence of many metastable phases with various reported compositions in the solid phase[31-33]. Especially in the middle concentration range, there exists various metastable solid phases. The reason for this may be attributed to the change in liquid state as a function of ethanol concentration and has been studied in detail by Koga and coworkers[34,35]. Additionally, high viscosity of the solutions at low temperature delays accomplishment of the solid-liquid equilibrium and facilitates formation of an amorphous state. A detailed solid – liquid phase diagram of water-ethanol is given by Takaizumi[33] from the freezing-thawing behavior of water-ethanol mixture studied using differential scanning calorimetric technique.

In the lower concentration region upto mole fraction $x_{EtOH} \sim 0.07$, the clathrate hydrate II $C_2H_5OH \cdot 17H_2O$ is easily formed and this has been confirmed by many authors. The mole fraction of $x_{EtOH} \sim 0.055$ corresponds to the composition, $C_2H_5OH \cdot 17H_2O$. In this region network of water exists with hydrogen bonding. In a relatively concentrated region, other types of hydrate start to co-exist; i.e., $C_2H_5OH \cdot 7.67H_2O$ corresponding to a mole fraction $x_{EtOH} \sim 0.11$ and $C_2H_5OH \cdot 5.67H_2O$ at a mole fraction range $x_{EtOH} \sim 0.15$. Takaizumi and coworkers[14] showed in a previous work that upto a concentration range $x_{EtOH} \sim 0.17$ ice Ih first freezes out from a supercooled solution, but beyond this concentration the first solid generated is not ice; rather ethanol hydrates are formed.

Although visualized through a number of experimental as well as simulation studies, the origin of anomalous behavior of water-ethanol binary mixture at low ethanol concentration is still not well understood. In order to understand the microscopic origin of the anomalies, we carry out temperature dependent study of water-ethanol binary mixture, particularly at low ethanol concentration. There are multitudes of competing interactions in this binary mixture, marked by hydrophobic interactions between ethyl-ethyl units, hydrogen bonding between water-water as well as ethanol-water molecules. In presence of such a wide range of competing interactions, we expect these mixtures to show interesting temperature dependent variations which can therefore be used to study relative importance of various interactions at different physical conditions favoring different micro-structural arrangements.

## II. Temperature dependent effect on local structure of aqueous ethanol solution

### a. Temperature dependence of radial distribution function (rdf) of ethyl groups

In order to explore the temperature dependent change in structural morphogenesis of ethanol molecules, we initially look into the temperature dependent variation of radial distribution function (rdf) of the ethyl groups at various concentrations of the mixture (Figure 1). We consider a dummy atom at the center of mass of the $CH_3$ and $CH_2$ groups, and calculate the rdf of those dummy atoms. Variation of rdfs gives a broad idea about the relative presence of other ethanol molecules in the neighboring shells. No appreciable change is observed in the first peak height of rdf with change in temperature at a lower concentration range of ethanol ($x_{EtOH}$ ~ 0.02-0.05). However, with gradual increase of ethanol concentration, lowering of temperature is found to have significant effect on rdfs of ethyl groups. This essentially means that with increase of ethanol concentration, lowering of temperature induces greater structural order in the system. Figure 1(f) provides a clear overview of the phenomenon where we plot the change in first peak height of rdf a function of temperature with increasing ethanol concentration.

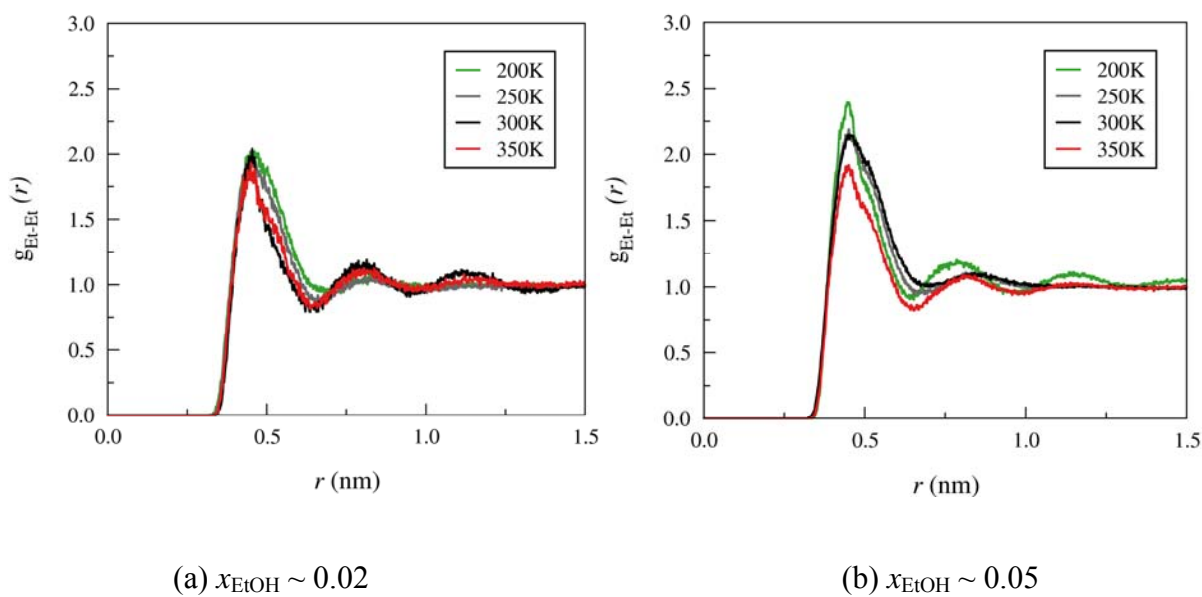

(a) $x_{EtOH}$ ~ 0.02     (b) $x_{EtOH}$ ~ 0.05

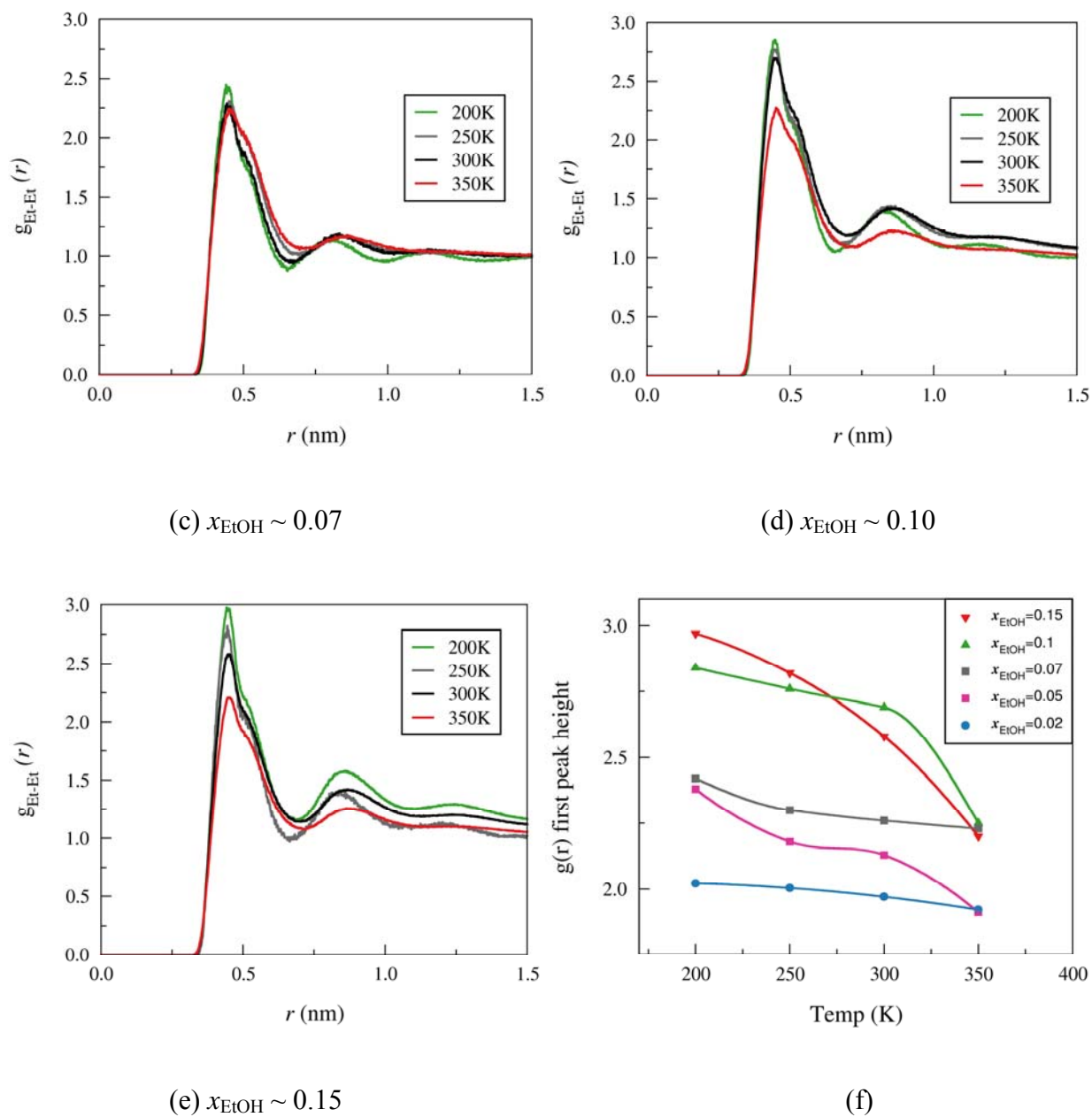

(c) $x_{EtOH} \sim 0.07$  (d) $x_{EtOH} \sim 0.10$

(e) $x_{EtOH} \sim 0.15$  (f)

**Figure 1. (a)-(e):** Plot of radial distribution function $g_{Et-Et}(r)$ between the ethyl groups of EtOH molecules as a function of temperature. The concentrations studied are $x_{EtOH} \sim 0.02; 0.05; 0.07; 0.1; 0.15$. **(f):** Change of first peak height of $g_{Et-Et}(r)$ with change of temperature.

**b.  Clustering of ethanol molecules: Effect of temperature**

In this section we analyze the microstructure of ethanol molecules formed at a low concentration of ethanol. At such low concentration as studied here, water molecules are expected to maintain a connected percolating cluster and ethanol molecules dispersed in the solution.

In order to understand the nature of micro-aggregation present in such binary systems, it is essential to observe the propensity of cluster formation. In fact, formation of micro-clusters in heterogeneous systems is known to exhibit strong temperature dependence. It has been already explored that ethanol molecules form spanning clusters and thereby show signature of percolation transition at a concentration regime $x_{EtOH}$~0.05-0.1[30]. We therefore intend to see how change of temperature affects formation of spanning clusters as well as the critical concentration range at which clusters start forming (percolation threshold).

The clusters of ethanol molecules are considered as a network formed via hydrophobic ethyl groups. We consider dummy atoms at the center of mass of $CH_3$ and $CH_2$ groups of ethanol. These dummy atoms serve as the building blocks of the network. Under the purview of percolation theory, a cluster is defined as a group of nearest neighboring occupied sites. In water-ethanol binary mixture, an estimate of the nearest neighboring shell of the dummy atoms (center of mass of ethyl groups) is obtained from the first minimum of their rdf as 0.65 nm. Therefore, we define that if the center of mass of the ethyl groups (i.e. the dummy atoms) are within a distance of 0.65 nm, then the corresponding ethanol molecules belong to the same cluster.

To check the formation of microclusters of ethanol as well as the corresponding temperature dependent effect, we look at the distribution of the clusters, given by $\langle sn_s \rangle$ – where $n_s$ is number of s-sized clusters present in the system, scaled by the total number of sites. Note that $\langle sn_s \rangle$ gives the probability density of finding a cluster of size *s*. In Figure 2, we plot the corresponding cluster size distribution demonstrating the distribution of $\langle sn_s \rangle$ along with change of temperature for different ethanol concentrations. We observe that at low concentration ($x_{EtOH}$~0.02-0.05), smaller sized clusters are prevalent in the system and no significant larger cluster is formed even at low temperature range. Larger clusters start appearing in the system at ethanol concentration $x_{EtOH}$~0.07, particularly at a lower temperature (200K) (Fig. 2(c)). At $x_{EtOH}$~0.1, a beautiful effect of temperature dependence on cluster size distribution is observed (Figure 2(d)). At 300K, larger sized clusters co-exist with smaller sized ones with comparable probability. However, on decreasing the temperature the smaller sized clusters gradually disappear and a continuous large cluster predominates. At $x_{EtOH}$~0.15, even at high temperature, continuous large cluster appears in the system. However the probability density for formation of large cluster increases markedly with decrease of temperature. This implies that although temperature does not have any significant effect on percolation transition threshold (which is a geometric transition), once percolation threshold is reached there is considerable temperature dependent effect on cluster formation. The temperature dependent effect essentially indicates that once percolation threshold is reached, greater structural order is induced in the system with subsequent lowering of temperature.

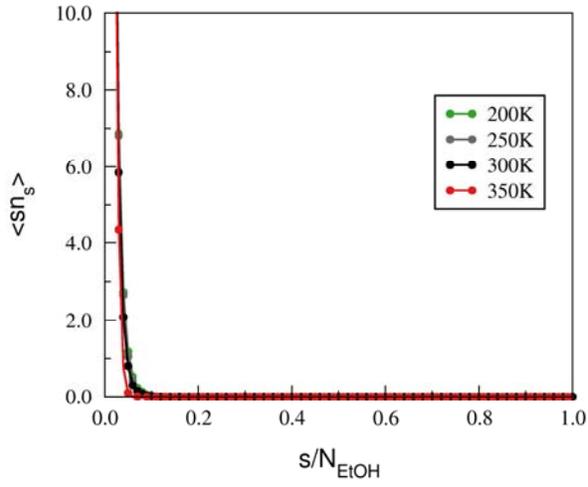

(a) $x_{EtOH} \sim 0.02$

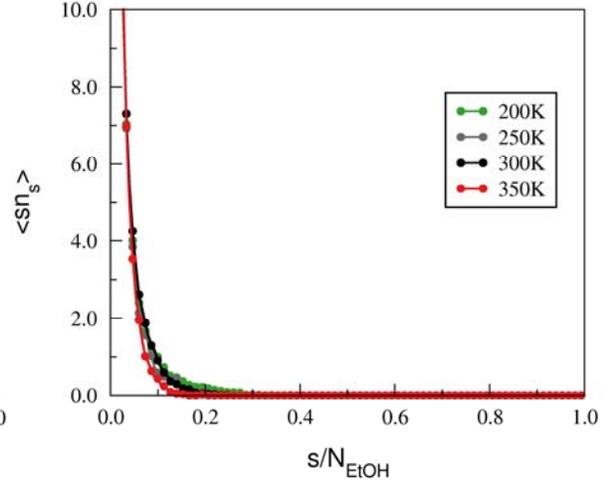

(b) $x_{EtOH} \sim 0.05$

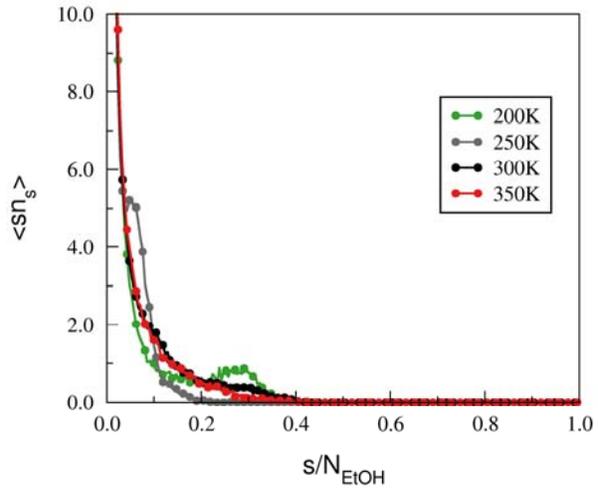

(c) $x_{EtOH} \sim 0.07$

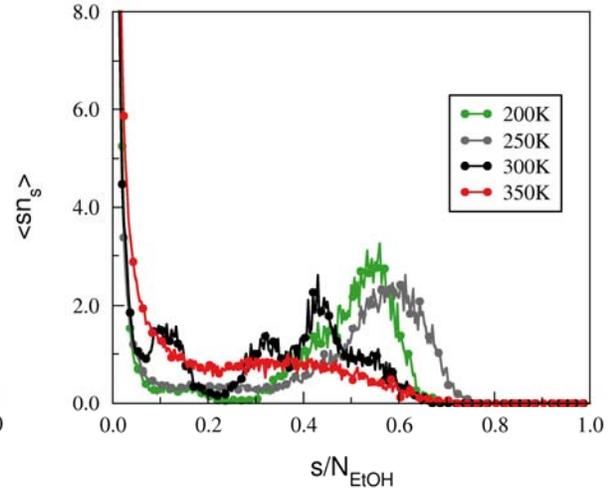

(d) $x_{EtOH} \sim 0.1$

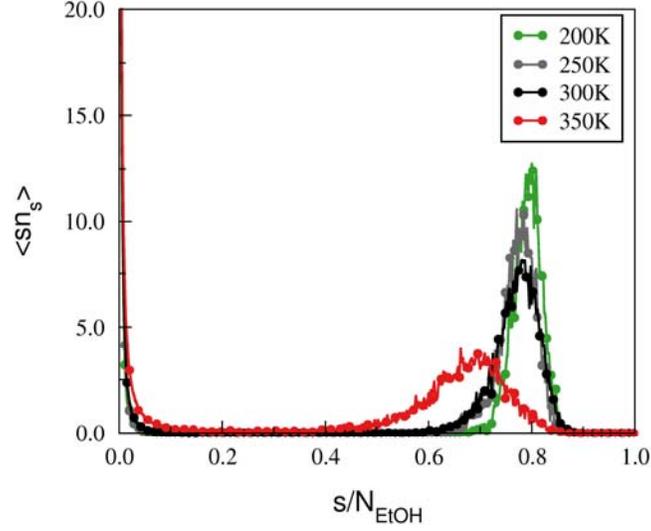

(e) $x_{EtOH} \sim 0.15$

**Figure 2.** Plot of $sn_s$ against $s/N_{EtOH}$ at various concentrations of water-ethanol binary mixture. The size of the cluster is $s$ and $n_s$ is the average number of s-sized clusters scaled by total number of ethanols $N_{EtOH}$.

c. **Effect of temperature on fractal dimension**

In case of percolation transition, the largest cluster is known to exhibit a fractal behavior at the percolation threshold $p_c$, following the asymptotic power law

$$s(p_c) \propto R_s^{d_f} \qquad (1)$$

The value of the universal exponent (in this case, fractal dimension $d_f$) in three dimensional systems is found to be $d_f^{3D} = 2.53$ [36]. The idea of fractal dimension implemented to describe shape of spanning clusters becomes clearer from the following statement made by Oleinikova and co-workers[37] as "the largest cluster of a system is a fractal object above the percolation threshold and no objects with fractal dimension lower than 2.53 can be infinite in three-dimensional space. Hence, the true percolation threshold is located where the fractal dimension of the largest cluster in the system reaches the critical value of 2.53." The statement makes it apparent that fractal dimension is a universal exponent and is dependent only on the dimension of the system. Here, we use the "sandbox method" to find the fractal dimension of ethanol clusters[38-40]. The spanning, largest cluster generated at and beyond percolation threshold is largely characterized by its shape. The more compact is the shape of the cluster higher is the value of $d_f$. We show the variation of cumulative radial distribution on $m(r)$ of the largest cluster of ethanol with radius $r$ in Figure 3. The cumulative radial distribution function is related to fractal dimension of the cluster by the following relation,

$$m(r) \propto r^{d_f} \qquad (2)$$

We obtain fractal dimensions of the largest ethanol clusters at different concentrations by fitting Equation (2) to cumulative radial distribution functions of Figure 3. We find that the shape of the largest cluster changes considerably with the lowering of temperature below the criticial concentration of percolation threshold. Percolation threshold appears at an ethanol mole fraction $x_{EtOH}$~0.1 marked by the critical value of fractal dimension ($d_f \approx 2.53$). Once the percolation threshold is reached, shape of the largest cluster is marginally affected with temperature change (as seen from Figure 3(e) at $x_{EtOH}$~0.15).

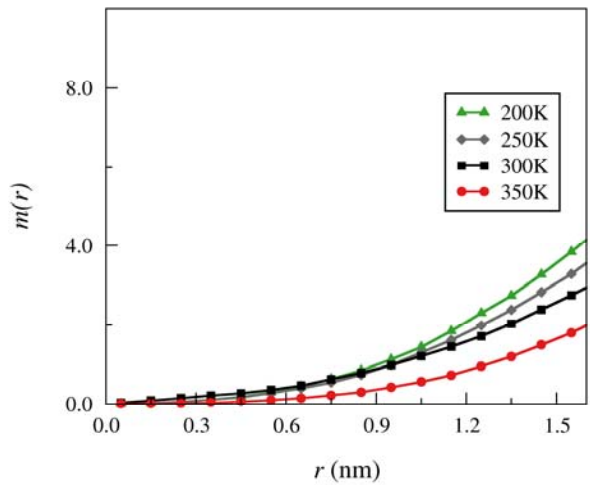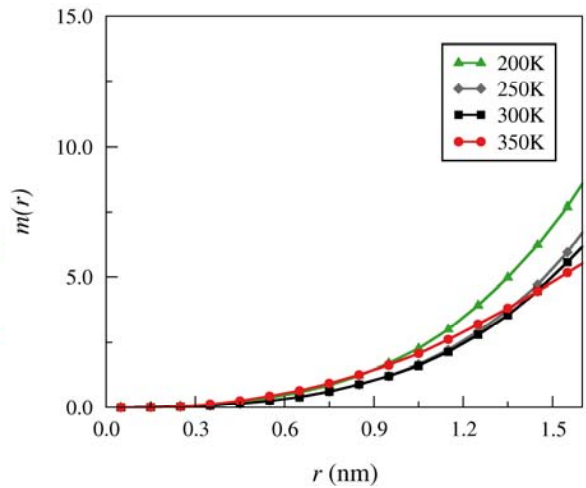

(a) $x_{EtOH} \sim 0.05$  (b) $x_{EtOH} \sim 0.07$

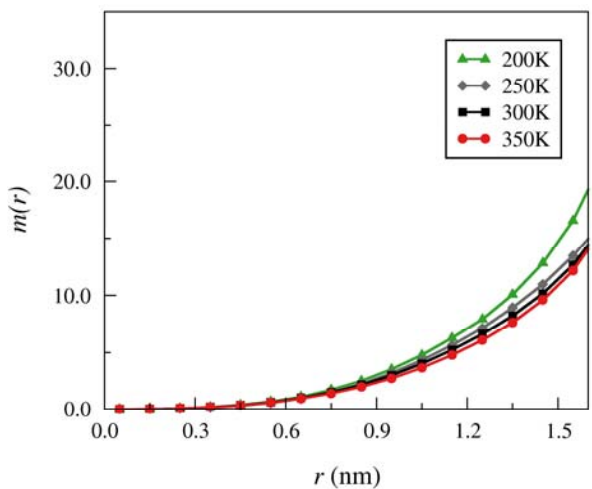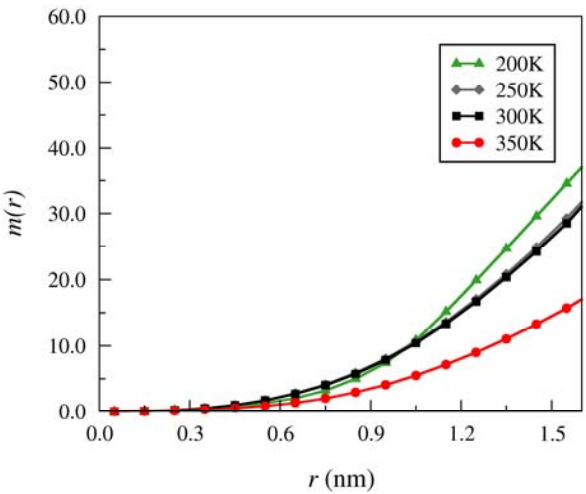

(c) $x_{EtOH} \sim 0.1$  (d) $x_{EtOH} \sim 0.15$

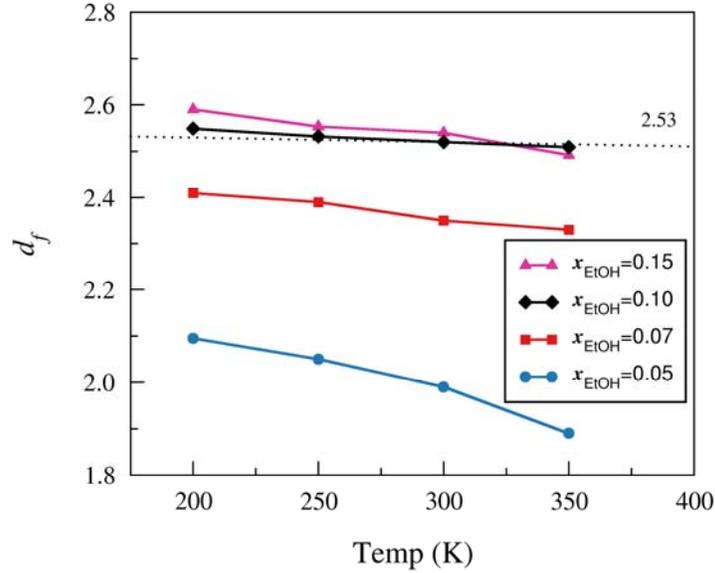

(e) Fractal dimension $d_f$ of the largest cluster of EtOH

**Figure 3. (a)-(d): Cumulative radial distribution *m(r)* of the largest ethanol cluster plotted against the radius r as a function of temperature. (e): fractal dimension $d_f$ of the largest cluster of ethanol (dashed line shows universal value of $d_f$ in 3D).**

### d. Temperature dependence of microscopic water structure: Tetrahedral order parameter

The microscopic structure of water in binary mixture of different cosolvents has been widely studied. However, the understanding is still far from being coherent. It can be apprehended that the formation of spanning cluster of any cosolvent will largely affect the tetrahedral ordered network of water. To understand the effect of ethanol self-aggregation (or clustering) as well as effect of temperature on the microscopic structure of water, we calculate the tetrahedral order parameter for water. The tetrahedral order parameter $t_h$ is defined as follows,

$$t_h = \frac{1}{n_{water}} \sum_k \left(1 - \frac{3}{8} \sum_{i=1}^{3} \sum_{j=i+1}^{4} \left[\cos\psi_{ikj} + \frac{1}{3}\right]^2\right) \quad (3)$$

Tetrahedral order parameter essentially measures the extent of tetrahedral arrangement maintained by the water molecules. A completely ordered tetrahedral network has a $t_h$ value of 1. The more ordered is the water structure, the higher is the $t_h$ value, whereas, the value decreases progressively with the extent of disorder introduced in the network. We plot the distribution of $t_h$ in Figure 4. We observe that the water structure is significantly perturbed even at a low concentration of ethanol. For $x_{EtOH}$~0.05 at 300K (Figure 4(a)), a broad distribution of $t_h$ is observed, giving an average value of 0.58. With decreasing temperature, the distribution moves towards a higher $t_h$ value signifying enhanced tetrahedral order introduced in the system. However, with increase of ethanol concentration and lowering of temperature a second small peak appears at lower $t_h$ value. This implies that along with tetrahedral order created in water structure with decrease of temperature, a significant part of the structure remains disordered as a consequence of formation of spanning ethanol clusters in the system.

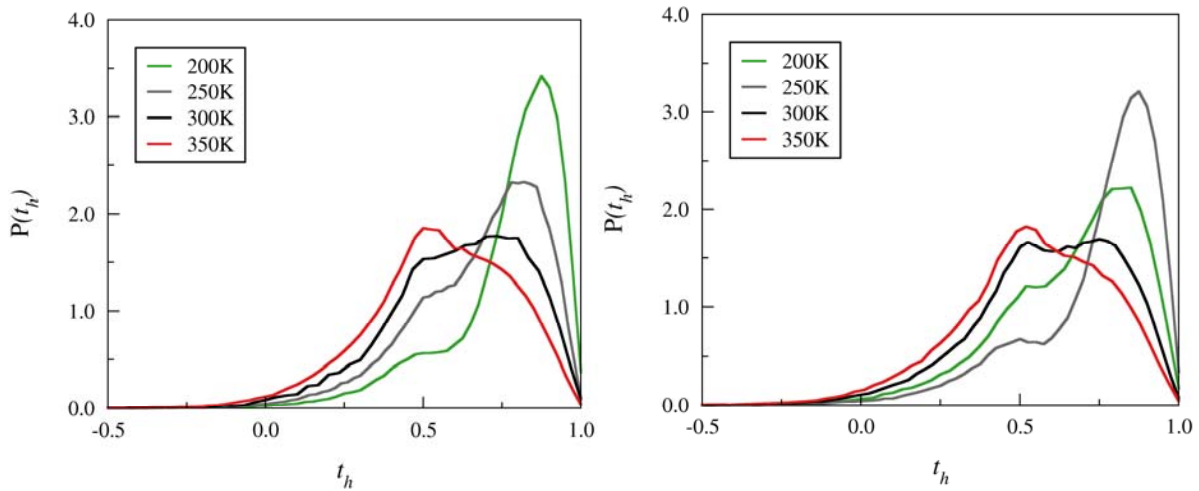

(a) $x_{EtOH}$ ~ 0.05     (b) $x_{EtOH}$ ~ 0.07

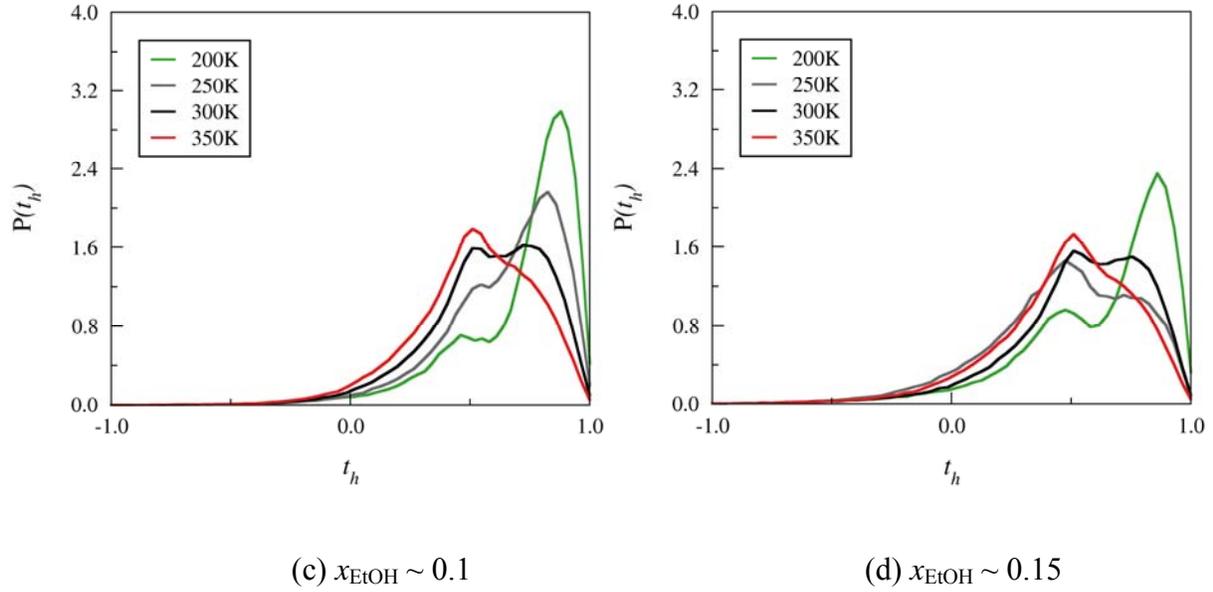

(c) $x_{EtOH} \sim 0.1$     (d) $x_{EtOH} \sim 0.15$

**Figure 4. (a)-(d): Distribution of tetrahedral order parameter $t_h$ of water with the change of temperature at different EtOH concentrations.**

We also plot the distribution of angle $\psi_{ikj}$ as a function of temperature for different ethanol concentrations in Figure 5. $\psi_{ikj}$ is the angle formed between the oxygen atoms of the $k$-th water molecule and the oxygen atoms of the nearest neighbors, $i$ and $j$. In this case also we find similar signature of disordered tetrahedral network with increasing ethanol concentration. The peak appearing at lower angular value of ~60° (arising due to interstitial water molecules) becomes progressively more prominent even at lower temperature with increasing ethanol concentration. This implies that there is enhancement of disorder in water structure with higher ethanol concentration that cannot be counterbalanced satisfactorily with lowering of temperature.

Until now we have looked into the formation of microstructures and effect of temperature on such structural arrangements in water-ethanol binary mixture. We find that both the microstructures of ethanol and water are significantly affected by temperature change. Next, we focus on the study of dynamical transition of the system with decreasing temperature.

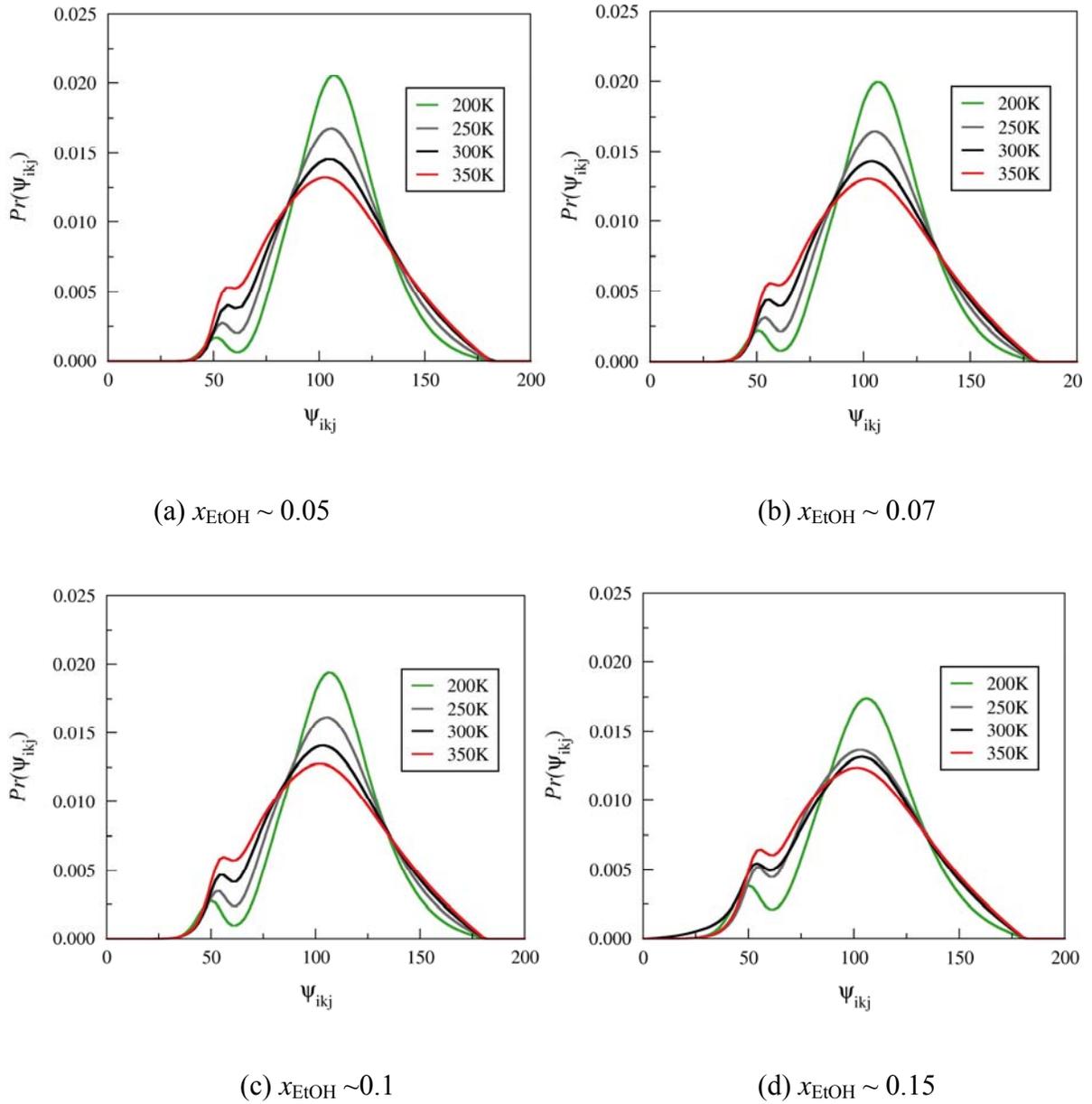

(a) $x_{EtOH} \sim 0.05$

(b) $x_{EtOH} \sim 0.07$

(c) $x_{EtOH} \sim 0.1$

(d) $x_{EtOH} \sim 0.15$

**Figure 5. Distribution of O—O—O angles ($\psi_{ikj}$) of water molecules at different concentrations of the binary mixture as a function of temperature. Water structure is found to be progressively perturbed with increasing EtOH concentration even at lower temperature regime.**

# III. Dynamical behavior of water-ethanol binary mixture: Temperature dependent effects

## a. Diffusion coefficient of water

In order to look into the change in dynamical behavior of the system, we evaluate the self-diffusion coefficient of water molecules along with change in temperature at different ethanol concentrations. The values of diffusion coefficient at different temperature and concentration regime are tabulated in Table 1.

**Table 1. Diffusion coefficient of water with increasing concentration of EtOH as a function of temperature**

| Diffusion Co-efficient | | | | | |
|---|---|---|---|---|---|
| $x_{EtOH}$ | **0.0** | **0.05** | **0.07** | **0.1** | **0.15** |
| **Temp (K)** | | | | | |
| **350** | 5.92 | 5.46 | 5.15 | 4.88 | 4.63 |
| **300** | 2.28 | 2.21 | 2.11 | 2.01 | 1.81 |
| **250** | 0.51 | 0.41 | 0.38 | 0.35 | 0.33 |
| **200** | 0.005 | 0.004 | 0.0037 | 0.0032 | 0.033 |

**Table 2. Glass transition temperature $T_0$ as obtained from Vogel-Fulcher-Tammann(VFT) equation fit of diffusion coefficients of water in water-EtOH binary mixture**

| $x_{EtOH}$ | $T_0$ |
|---|---|
| 0.0 | 166.2 |
| 0.05 | 150.6 |
| 0.07 | 149.9 |
| 0.10 | 148.1 |
| 0.15 | 143.4 |

We find that in case of water-ethanol binary mixture system, change in the value of diffusion coefficient of water with increase of ethanol concentration is reasonably significant. We fit the temperature dependent diffusion coefficient values to the empirical Vogel-Fulcher-Tammann(VFT) equation, according to the following expression to find the glass transition temperature. It is well-known that glass forming liquids show markedly non-Arrhenius behavior as they are supercooled the below freezing point. SPC/E water the glass transition temperature is known to appear around 165K. The temperature dependence of this non-Arrhenius behavior is often well represented by VFT equation, given by the following expression,

$$D = D_0 \exp\left(\frac{-E}{T - T_0}\right) \quad (4)$$

Here $D$ is the temperature dependent diffusion coefficient, $T_0$ is often related to glass transition temperature, $D_0$ and $E$ are fitting constants. The VFT fit of diffusion coefficient is presented in Figure 6. We obtain $T_0$ for each ethanol concentration. The data is presented in the following table (Table 2). Interestingly, the predicted glass transition temperature for different concentrations of aqueous ethanol solution is not found to be much deviated from that for the bulk water. This essentially suggests that the dynamical behavior of water is not appreciably affected even at a moderate ethanol concentration. In search of further consolidated evidence of this fact, we explore the dynamic heterogeneity of the system in presence of ethanol.

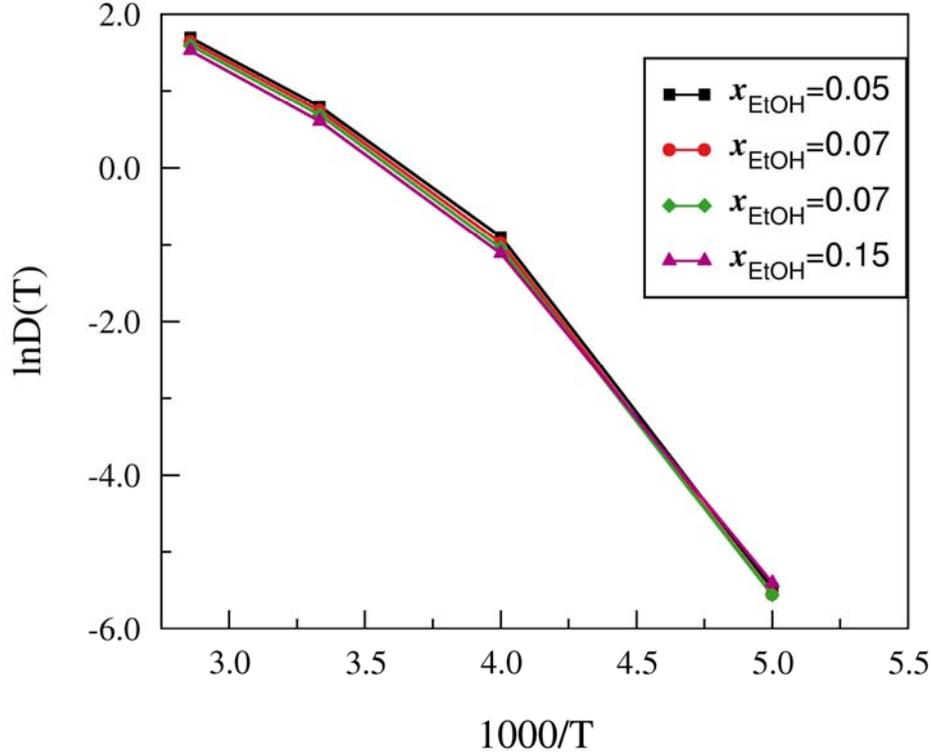

**Figure 6. Vogel-Fulcher-Tammann (VFT) equation fit of diffusion coefficient of water as a function of temperature with increasing EtOH concentration.**

**b.     Temperature dependent effect of static and dynamic heterogeneity: Calculation of non-Gaussian order parameter and non-linear response function $\chi_4(t)$**

The presence of microscopic heterogeneity in any complex system is known to be reflected in their substantial non-Gaussian behavior. The most frequently used indicator of non-Gaussian behavior is the parameter $\alpha_2(t)$ [41], which entails a ratio of the second and fourth moments of the displacement distribution. $\alpha_2(t)$ is defined by equation (5) as,

$$\alpha_2(t) = \frac{\langle \Delta r^4(t) \rangle}{\left(1+\frac{2}{d}\right)\langle \Delta r^2(t) \rangle^2} - 1 \quad (5)$$

Where $d=3$ for three dimensional systems. $\alpha_2(t)$ is defined to be zero when the distribution of displacements is Gaussian. Therefore, it goes to zero in the long time. It can also be shown that $\alpha_2(t)$ goes to zero at very short times when the motion is ballistic. In the intermediate times, $\alpha_2(t)$ becomes non-zero as different molecules in different regions diffuse at different speeds, thus making the distribution of displacement non-Gaussian. $\alpha_2(t)$ is often referred to as a measure of static heterogeneity because displacements of different molecules can remain different from each other which can be captured easily by plotting $\alpha_2(t)$ against time[42]. When some regions are liquid-like and some solid-like, the function does not go to zero within the time scale of simulations, although displays a gradual decrease. As $\alpha_2(t)$ is calculated over all the particles of the system, long regions display different relaxation times. That is the reason why $\alpha_2(t)$ is expected to probe static heterogeneity present in the system.

We present the plots of $\alpha_2(t)$ of water for three different temperatures (350K, 300K and 250K) at different ethanol concentrations in Figure 7. Since the dynamics of $\alpha_2(t)$ is very slow at 200K even for neat bulk water, we do not take that temperature into account in this set of calculations. Interestingly, we find that $\alpha_2(t)$ of water does not show significant deviation from the corresponding behavior of bulk water even at a comparatively higher ethanol concentration. The peak position shifts marginally to a longer time scale with increasing ethanol concentration and lowering of temperature. The minor change in $\alpha_2(t)$ implies the relatively weak presence of large scale static inhomogeneity in the system.

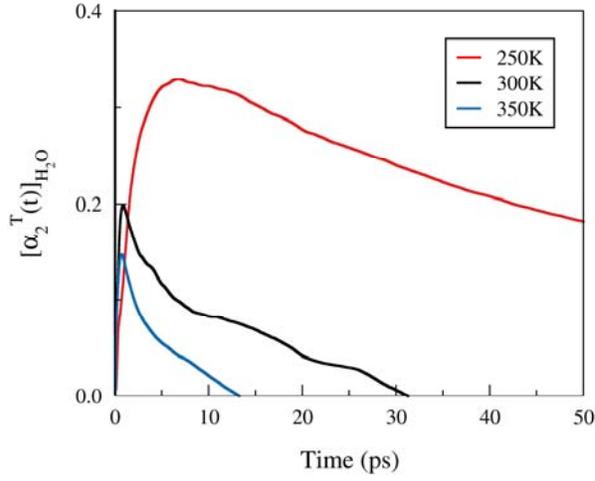

(a) $x_{EtOH} \sim 0.0$ (water)

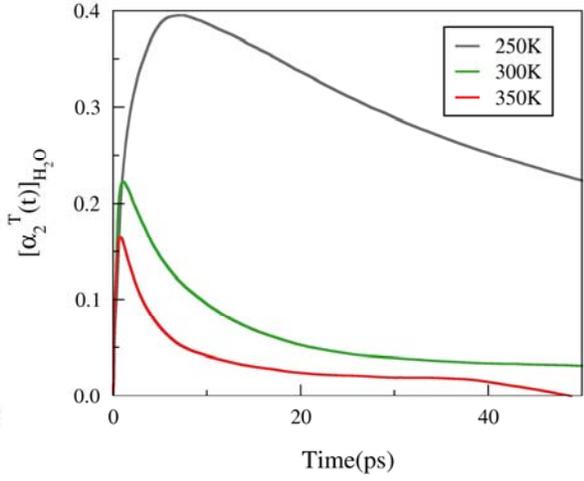

(b) $x_{EtOH} \sim 0.07$

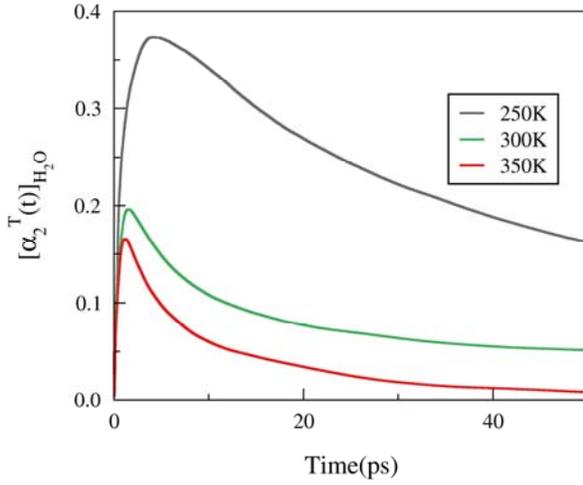

(c) $x_{EtOH} \sim 0.1$

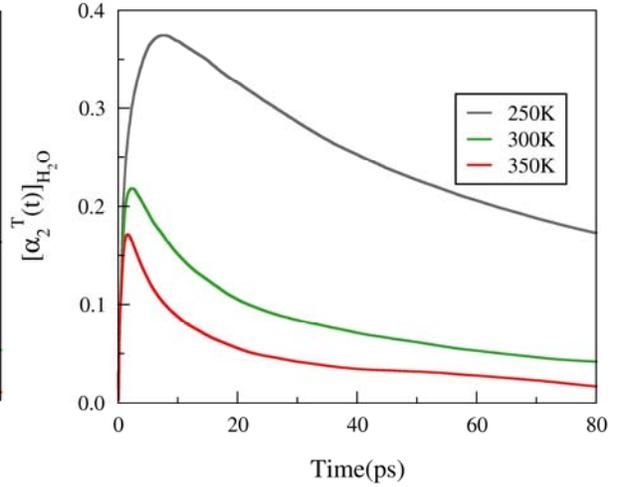

(d) $x_{EtOH} \sim 0.15$

**Figure 7. Non-Gaussian order parameter $\alpha_2(t)$ of water as a function of temperature with increasing ethanol concentration. $\alpha_2(t)$ of water is found to be marginally affected compared to that of bulk water with increasing ethanol concentration.**

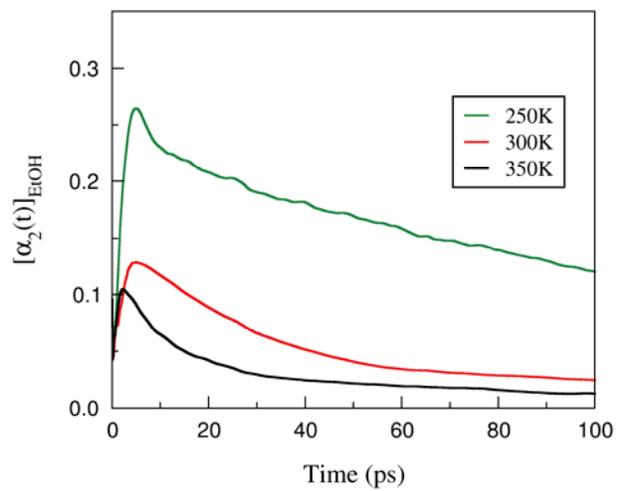

(a) $x_{EtOH} \sim 0.02$

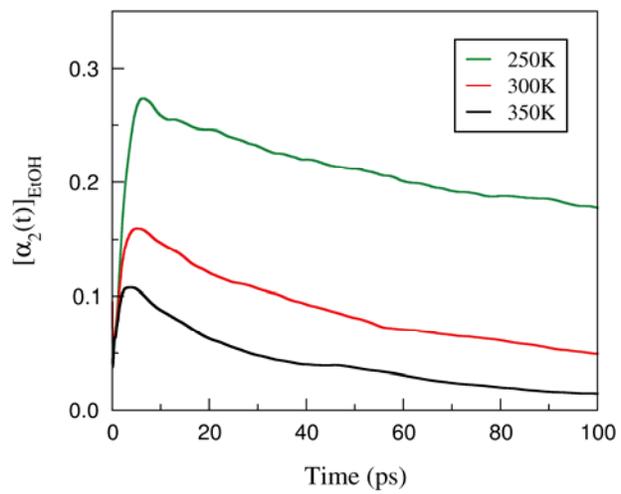

(b) $x_{EtOH} \sim 0.05$

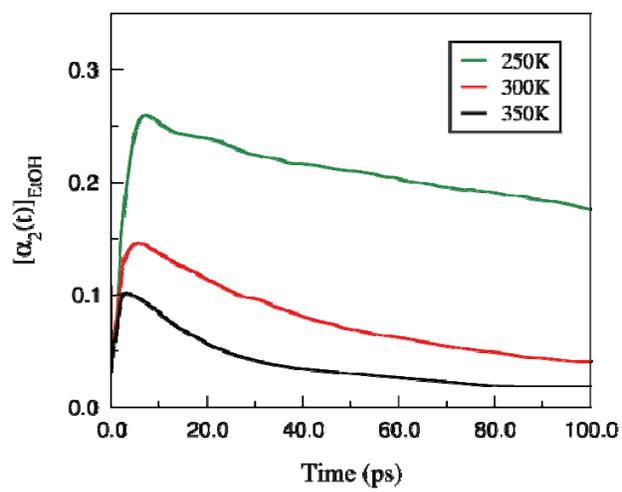

(c) $x_{EtOH} \sim 0.07$

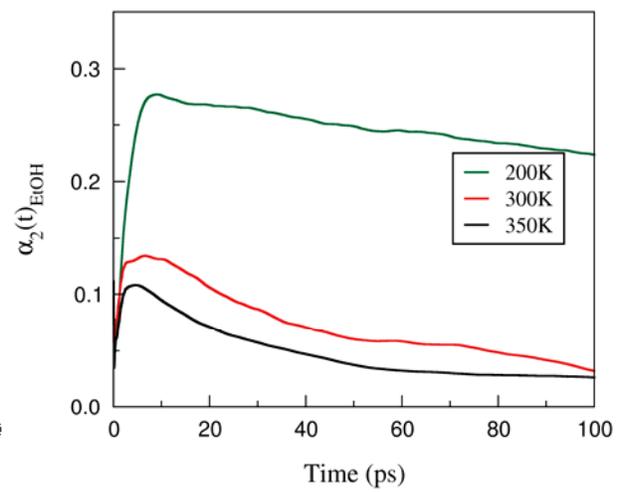

(d) $x_{EtOH} \sim 0.1$

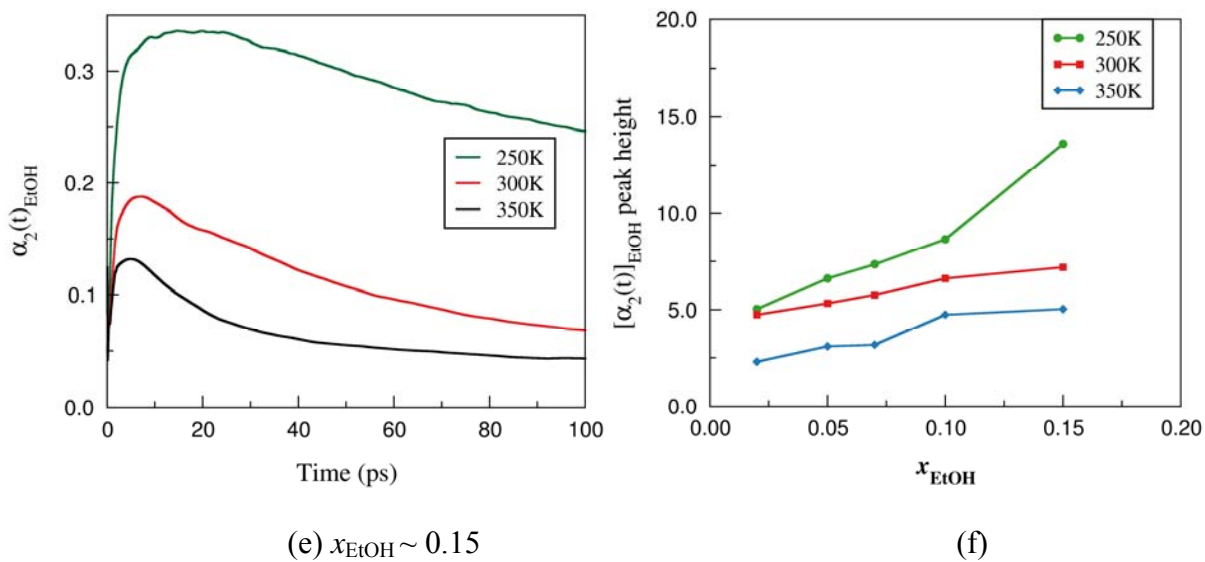

(e) $x_{EtOH} \sim 0.15$          (f)

**Figure 8. (a)-(e): Non-Gaussian order parameter $\alpha_2(t)$ of ethanol as a function of temperature plotted at different ethanol concentrations. $\alpha_2(t)$ of ethanol is found to be somewhat affected with increasing ethanol concentration as well as lowering of temperature. (f) Plot of peak position of $\alpha_2(t)$ (in ps) as a function of ethanol composition at three different temperatures.**

In order to understand the dynamical behavior of ethanol molecules in this class of binary mixtures, we calculate the $\alpha_2(t)$ of ethanol molecules as a function of temperature with increasing ethanol concentration (Figure 8). In this case, we follow the dynamics of the central carbon atom of ethanol molecule which is expected to reveal the overall dynamical behavior of the molecule itself. As can be anticipated, decay of $\alpha_2(t)$ of ethanol is found to be considerably slower than that of water (greater than 100 ps) even at a temperature as high as 350K. The corresponding explanation for slower decay of $\alpha_2(t)$ can be conveniently attributed to the bigger size of ethanol molecules. However, the timescale of inhomogeneity is found to shift marginally to a higher value with increasing ethanol concentration. This change in the trend of $\alpha_2(t)$ is found to be similar to that of water.

To get a further insight into the temperature and concentration dependence of dynamical behavior of the system, we calculate the non-linear response function $\chi_4(t)$ which measures the lengthscale of dynamical heterogeneity (as explained below) present in the system[43,44]. $\chi_4(t)$, also known as four point susceptibility, is related to four point density correlation function $G_4$ by the following relation,

$$\chi_4(t) = \frac{\beta V}{N^2} \int dr_1 dr_2 dr_3 dr_4 w\left(|r_3 - r_4|\right) \times G_4\left(r_1, r_2, r_3, r_4, t\right) \tag{6}$$

The four point density correlation function $G_4$ can be written as,

$$\begin{aligned}G_4(r_1,r_2,r_3,r_4,t) &= \langle \rho(r_1,0)\rho(r_2,t)\rho(r_3,0)\rho(r_4,t)\rangle \\ &\quad - \langle \rho(r_1,0)\rho(r_2,t)\rangle \times \langle \rho(r_3,0)\rho(r_4,t)\rangle\end{aligned} \tag{7}$$

Which essentially implies that $\chi_4(t)$ is dominated by a range of spatial correlation between the localized particles in the fluid. It can be shown that expression of $\chi_4(t)$ is also equivalent to the following relation,

$$\chi_4(t) = \frac{\beta}{\rho N}\left[\langle Q^2(t)\rangle - \langle Q(t)\rangle^2\right] \tag{8}$$

Where $Q(t)$ is a time dependent order parameter which measures the localization of particles around a central molecule through an overlap function which is unity inside a region $a$ and $0$ otherwise. . Thus, $\chi_4(t)$ may remain unity for a longer time in a slow, solid-like region but goes to zero as long as a transition to liquid-like region occurs because the tagged atom or molecule can now move beyond the specified distance measure $a$. It is important to realize that the $Q(t)$ is a sum over all the atoms and molecules of the system. Therefore, the time dependent distribution of $Q(t)$, $P(Q,t)$ that is required to obtain $\chi_4(t)$ as defined by Equation 8, is obtained from many trajectories of the entire system because at a given time, a system has a given value of $Q(t)$. $\chi_4(t)$ therefore provides a measure of the collective dynamical state of the system, and somewhat different from the non-Gaussian order parameter $\alpha_2(t)$.

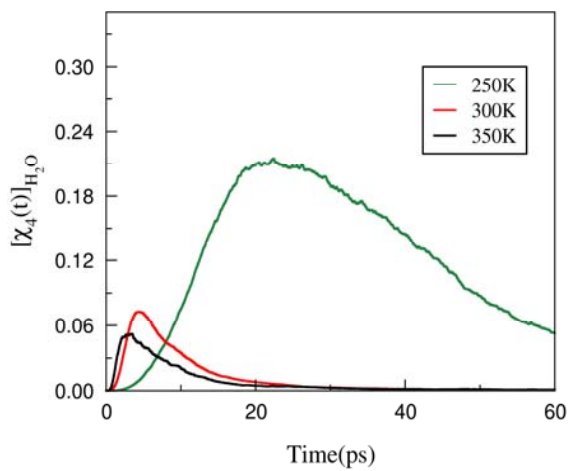

(a) Bulk water

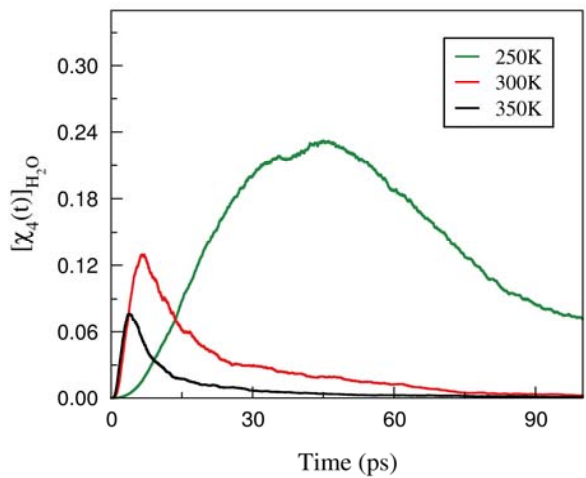

(b) $x_{EtOH} \sim 0.07$

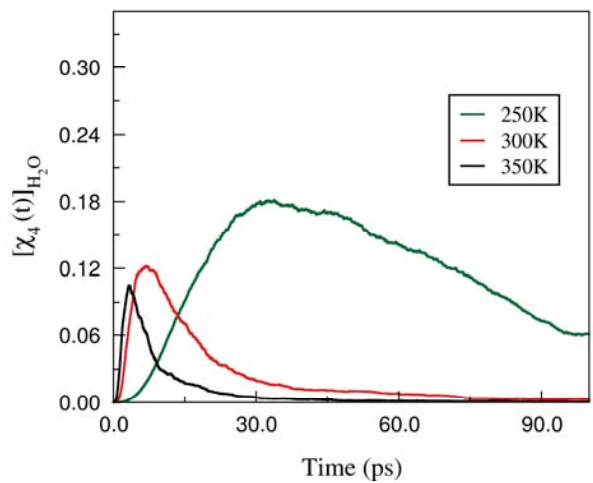

(c) $x_{EtOH} \sim 0.1$

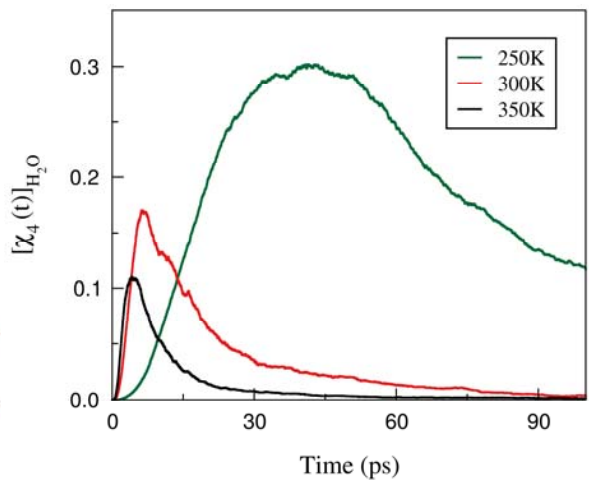

(d) $x_{EtOH} \sim 0.15$

**Figure 9. Time dependence of non-linear dynamic response function $\chi_4(t)$ at three different temperatures with increasing ethanol concentration.**

In Figure 9, we demonstrate the non-linear response function, $\chi_4(t)$, of water molecules (computer from oxygen atom displacements) for three different temperatures (350K, 300K and 250K) with increasing ethanol concentration. In this case also, the shift of peak position to a longer time scale is found to be relatively insignificant. To further explore the fact, we plot the non-linear response function $\chi_4(t)$ of ethanol molecules over the entire low concentration regime (Figure 10).

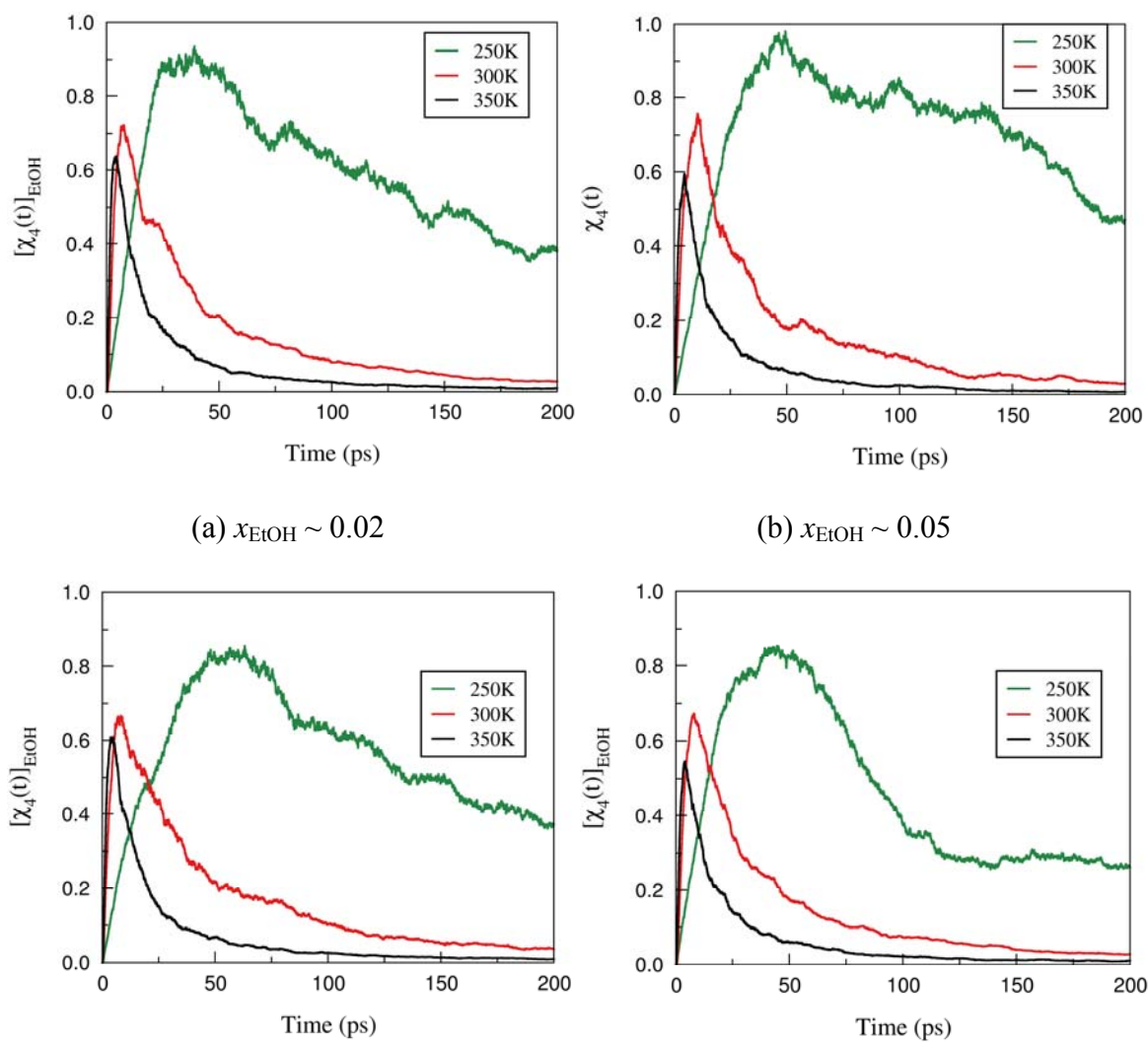

(a) $x_{EtOH} \sim 0.02$  (b) $x_{EtOH} \sim 0.05$

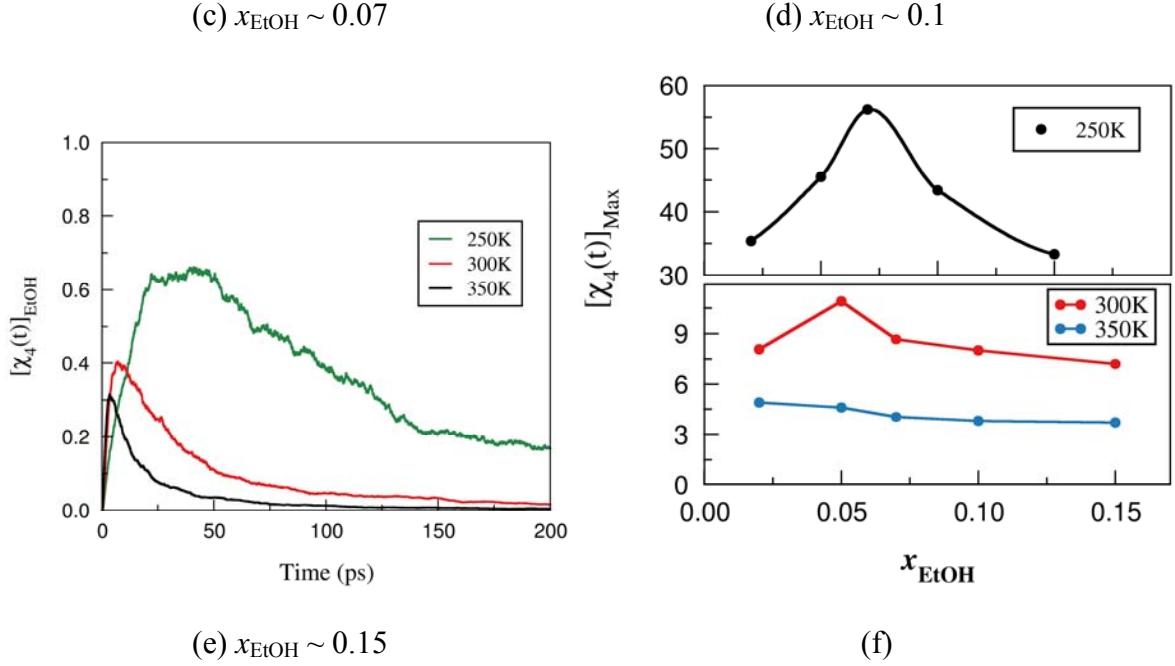

(c) $x_{EtOH} \sim 0.07$  (d) $x_{EtOH} \sim 0.1$

(e) $x_{EtOH} \sim 0.15$  (f)

**Figure 10. (a)-(e): Non-linear dynamic response function $\chi_4(t)$ of ethanol molecules at different ethanol concentrations as a function of temperature. (f) Plot of peak position of $\chi_4(t)$ (in ps) as a function of ethanol concentration. Signature of anomalous dynamics is found from the plot of $\chi_4(t)$ peak position at the concentration range $x_{EtOH} \sim 0.05$-0.1, particularly at low temperature.**

This set of plot reveals a number of interesting features. $\chi_4(t)$ is found to capture the anomalous dynamic heterogeneity of ethanol molecules rather faithfully. It is found that the dynamic heterogeneity is particularly pronounced at around ethanol concentration range $x_{EtOH} \sim 0.05$-0.1 at 250K. Even at 300K, the signature of anomalous dynamic heterogeneity is reasonably significant at this concentration range. We discuss the plausible explanations in detail in the next section.

## IV.  Discussion and concluding remarks

The well known anomalies in water-ethanol binary mixture at low concentration range have been addressed for a long time and several experimental and theoretical studies (discussed in detail in the Introduction section) have been carried out to understand the molecular origin. The plausible origin of such anomalous behavior is attributed to the percolation driven structural aggregation of the ethanol molecules leading to formation of micro-segregated phases. In order to obtain a complete view of the intermolecular interactions responsible for such structural transformation as well as understand the stability of structures formed and nature of altered dynamics, we have carried out temperature dependent study of water-ethanol binary mixture at low ethanol concentration, particularly below and above the critical percolation transition region ($x_{EtOH}$~0.05-0.1). We find that below the percolation threshold the structural arrangement of the system and precisely that of ethanol molecules is largely unaffected with the lowering of temperature. However, at and beyond the percolation threshold, we observe temperature lowering induces enhanced structural order in the system that is reflected in the $g_{Et-Et}(r)$ and cluster size distribution.

We have also looked into the water structure and the extent of disorder introduced due to formation of spanning clusters of ethanol. We find that the water structure is considerably disrupted even at low ethanol concentration that is visible from the distribution of tetrahedral order parameter.

Next, we explore temperature dependence of the dynamical behavior of the system. We calculate diffusion coefficient of water molecules and fit the temperature dependent value of diffusion coefficients to the VFT equation in order to find the possible glass transition at low temperatures. The glass transition temperature did not change appreciably from that of the pure water. Next we calculate non-Gaussian order parameter ($\alpha_2(t)$) which shows shift of the peak to a relatively longer timescale with increasing ethanol concentration and decreasing temperature. This essentially signifies presence of static heterogeneity in the system, although the shift is found to be relatively weak.

In contrast, the non-linear response function $\chi_4(t)$ of ethanol molecules demonstrates signatures of dynamic heterogeneity rather faithfully, particularly at low temperature and in the concentration range $x_{EtOH}$ ~ 0.05-0.1 (precisely the concentration regime where microscopic structural transformation takes place). Note that our earlier studies have shown that ethanol clusters in water-ethanol binary mixture have reasonably short life time, compared to otherwise similarly behaving binary mixtures such as, water-dimethyl sulfoxide (DMSO) or water-tertiary butyl alcohol (TBA). TBA clusters have lifetime in excess of 20 ps while ethanol clusters are found to have lifetime only of the order of a ps[45]. The short lifetime of ethanol clusters offers an explanation of the notable absence of static and dynamic heterogeneity in this system at relatively higher temperature. However, the dynamic heterogeneity of ethanol molecules is found to be quite pronounced when temperature is lowered which makes the lifetime of these clusters sufficiently elongated (Figure 10). It is interesting to note existence of anomalous dynamic heterogeneity even at 300K (Figure 10(f)). We observe that static heterogeneity also increases on lowering temperature (Figure 7-8). In fact, the weak signature of static and dynamic heterogeneity appearing for water molecules can be attributed to the transient nature of the ethanol clusters.

It is interesting to note the maximum in the timescale of $\chi_4(t)$ at ethanol mole fraction $x_{EtOH}$ ~ 0.05-0.07. This is precisely the concentration range where all other anomalies in water-ethanol binary mixture are observed. This further supports the idea of a weak structural transformation in the mixture's configuration space that indeed occurs in this range, but is reflected only in certain dynamic properties as the transformation is made weak by the ultra short lifetime of the ethanol clusters. The ultrafast timescale of ethanol clusters makes the detailed quantitative characterization of the complex behavior of this solution with different competing interactions a rather arduous task. Nevertheless, this work brings out the essential microscopic behavior of this well-known binary mixture in terms of structure and dynamics and reveals the temperature dependent behavior of the system that has not been anticipated before.

To summarize, water-ethanol binary mixture exhibits interesting temperature dependence even at low solute concentration that is manifested in the microscopic structural and dynamical behavior. All the analyses indicate percolation induced formation of micro-structure in the system at a critical ethanol concentration marked by percolation threshold that gets progressively more ordered with lowering of temperature. The presence of transient ethanol clusters at and beyond percolation threshold is reflected in the anomalous change in dynamic heterogeneity of the system. Finally we note that time scales of water-ethanol mixtures are such that they can influence ultrafast chemical processes such as solvation dynamics and charge transfer processes.

## V.     Simulation Details

We have simulated water-ethanol binary mixture at five different concentrations ($x_{EtOH}$ ~ 0.02, 0.05, 0.07, 0.1, 0.15) and four different temperatures (350K, 300K, 250K and 200K). The pressure has been kept at 1 Bar for all the simulations. The ethanol molecules are treated as united atoms in GROMOS53a6 force field[46]. Extended simple point charge model (SPC/E) is used for water[47]. To perform Molecular Dynamics simulations, we have used GROMACS (v4.5.5) which is highly scalable and efficient molecular simulation engine[48-51]. We have taken considerably large systems for each concentration (~4000 molecules altogether) to eliminate the finite size effect, if any, present in the system. The box size has been taken to be as large as ~ 8nm. After performing steepest descent energy minimization, equilibration of the system is done for 5 ns keeping temperature and volume constant. Followed by this, an equilibration is performed for 5 ns keeping pressure and temperature constant. Finally production run has been executed for 50 ns in a NPT ensemble. Temperature is kept constant using Nose-Hoover Thermostat[52,53] and Parinello-Rahman Barostat[54] is used for pressure coupling. Periodic boundary conditions are applied and non-bonded force calculations are employed by applying grid system for neighbor searching. The cut-off radius taken for neighbor list and van der Waals interaction was 1.4nm. To calculate electrostatic interactions we used Particle Mesh Ewald (PME)[55] with a grid spacing of 0.16nm and an interpolation order of 4.

## Acknowledgements:

This work was supported in parts by grant from DST, India. B Bagchi thanks JC Bose Fellowship (DST) for a partial support of the research.